\documentclass[twocolumn,usenames,dvipsnames]{aastex62}

\usepackage{xcolor}

\def\rev#1{#1} 

\definecolor{maroon}{cmyk}{0, 0.87, 0.68, 0.32}

\newcommand{\picwide}[2]{
	\begin{figure}
	 \includegraphics[width=\columnwidth]{#1}
	 \caption{#2}
	 \label{#1}
	\end{figure}}
\graphicspath{{./}{figures/}}

\accepted{March 2, 2020}
\submitjournal{ApJ}

\shorttitle{Proxima Centauri b}
\shortauthors{Scheucher et al.}

\begin{document}

\title{Proxima Centauri b : A Strong Case for including Cosmic-Ray-induced Chemistry in Atmospheric Biosignature Studies}

\correspondingauthor{Markus Scheucher}
\email{scheucher@tu-berlin.de, markus.scheucher@dlr.de}

\author[0000-0003-4331-2277]{M. Scheucher}
\affil{Zentrum f\"{u}r Astronomie und Astrophysik, Technische Universit\"{a}t Berlin, 10623 Berlin, Germany}
\affil{Institut f\"{u}r Planetenforschung, Deutsches Zentrum f\"{u}r Luft- und Raumfahrt, 12489 Berlin, Germany}

\author[0000-0001-5622-4829]{K. Herbst}
\affil{Institut f\"{u}r Experimentelle and Angewandte Physik, Christian-Albrechts-Universit\"at zu Kiel, 24118 Kiel, Germany}

\author{V. Schmidt}
\affil{Institut f\"{u}r Meteorologie und Klimaforschung, Karlsruher Institut f\"{u}r Technologie, 76344 Eggenstein-Leopoldshafen, Germany}

\author{J. L. Grenfell}
\affil{Institut f\"{u}r Planetenforschung, Deutsches Zentrum f\"{u}r Luft- und Raumfahrt, 12489 Berlin, Germany}

\author[0000-0001-7196-6599]{F. Schreier}
\affil{Institut f\"{u}r Methodik der Fernerkundung, Deutsches Zentrum f\"{u}r Luft- und Raumfahrt, 82234 Oberpfaffenhofen, Germany}

\author{S. Banjac}
\affil{Institut f\"{u}r Experimentelle and Angewandte Physik, Christian-Albrechts-Universit\"at zu Kiel, 24118 Kiel, Germany}

\author{B. Heber}
\affil{Institut f\"{u}r Experimentelle and Angewandte Physik, Christian-Albrechts-Universit\"at zu Kiel, 24118 Kiel, Germany}

\author{H. Rauer}
\affil{Zentrum f\"{u}r Astronomie und Astrophysik, Technische Universit\"{a}t Berlin, 10623 Berlin, Germany}
\affil{Institut f\"{u}r Planetenforschung, Deutsches Zentrum f\"{u}r Luft- und Raumfahrt, 12489 Berlin, Germany}
\affil{Institut f\"{u}r Geologische Wissenschaften, Freie Universit\"{a}t Berlin, 12249 Berlin, Germany}

\author{M. Sinnhuber}
\affil{Institut f\"{u}r Meteorologie und Klimaforschung, Karlsruher Institut f\"{u}r Technologie, 76344 Eggenstein-Leopoldshafen, Germany}



\begin{abstract}
Due to its Earth-like minimum mass of 1.27 M$_{\text{E}}$ and its close proximity to our Solar system, Proxima Centauri b is one of the most interesting exoplanets for habitability studies. Its host star, Proxima Centauri, is however a strongly flaring star, which is expected to provide a very hostile environment for potentially habitable planets. We perform a habitability study of Proxima Centauri b assuming an Earth-like atmosphere under high stellar particle bombardment, with a focus on spectral transmission features. We employ our extensive model suite calculating energy spectra of stellar particles, their journey through the planetary magnetosphere, ionosphere, and atmosphere, ultimately providing planetary climate and spectral characteristics, as outlined in \cite{Herbst-etal-2019b}. Our results suggest that together with the incident stellar energy flux, high particle influxes can lead to efficient heating of the planet well into temperate climates, by limiting CH$_4$ amounts, which would otherwise run into anti-greenhouse for such planets around M-stars. We identify some key spectral features relevant for future spectral observations: First, NO$_2$ becomes the major absorber in the visible, which greatly impacts the Rayleigh slope. Second, H$_2$O features can be masked by CH$_4$ (near infra-red) and CO$_2$ (mid to far infra-red), making them non-detectable in transmission. Third, O$_3$ is destroyed and instead HNO$_3$ features become clearly visible in the mid to far infra-red. Lastly, assuming a few percent of CO$_2$ in the atmosphere, CO$_2$ absorption at 5.3~$\mu$m becomes significant (for flare and non-flare cases), strongly overlapping with a flare related NO feature in Earth's atmosphere.

\end{abstract}



\section{Introduction} \label{sec:intro}
Given the recent, exciting discoveries of terrestrial-sized planets orbiting M-stars, together with the higher activity of many M-stars compared to our Sun \citep[e.g.][]{Reid2005,scalo2007}, a better understanding of the influence and impact of such active host stars upon planetary habitability is crucial for the search for extra-terrestrial life and improving our understanding of Earth-like planets. Proxima Centauri b (hereafter Prox Cen b) is one of the most interesting exoplanets to date in terms of studying potential habitability \citep[see, e.g.,][]{turbet2016,Dong2017,meadows2018,Berdyugina2019}. With a minimum of 1.27 Earth masses \citep{Anglada2016}, it may be similar in bulk properties to Earth. Although it receives only 65\% of the mean total stellar irradiation (TSI) compared to Earth, \cite{meadows2018} showed that an Earth-like atmosphere with, e.g. a surface carbon dioxide concentration of a few percent could lead to habitable conditions.\\
Proxima Centauri (hereafter Prox Cen), is an M5.5Ve flaring star. While direct observations of coronal mass ejections (CMEs) and corotating interaction regions (CIRs) are still challenging, model extrapolations from the Sun's flare-CME correlation can be used to estimate the bombardment by stellar energetic particles (SEP). While the planetary magnetosphere could shield the planet from the majority of low-energy SEPs, the multitude of flares - and possible CMEs - of active M-stars may cause long-lasting changes to the planet's atmospheric mass, composition and surface conditions \citep[e.g.][]{vidotto2013}. There is an ongoing debate as to whether such close-in planets orbiting active M-stars would be stripped of their atmospheres, e.g. if they lie within the star's Alfv\'en sphere, leaving them without magnetospheric protection \cite[e.g.][]{lammer2009,Airapetian2017}.\\
Including the impact of stellar high energetic particles in habitability studies in a self-consistent way requires a broad understanding of stellar, astrospheric, magnetospheric, and ion- and neutral chemical processes within the planet's atmosphere. Modeling efforts by e.g. \citet{segura2010}, \citet{grenfell2012}, \citet{tabataba2016}, \citet{Tilley2017}, and \citet{Scheucher2018} have parameterized the top-of-atmosphere (TOA) incoming particle energy distributions, secondary particle generation in air showers, ionization of the atmosphere, and its impact on neutral atmospheric composition, using different methods. \citet{Herbst-etal-2019b} took this one step further by coupling cosmic-ray-induced magnetospheric, ionospheric and lower atmospheric processes in an interactive model suite.\\
Such studies are crucial in order to understand the expected range of atmospheres of such planets lying in - or close to - the habitable zone, as well as to better understand and interpret atmospheric spectra of next generation space missions such as JWST, HabEX, and LUVOIR, plus ground-based telescopes like the ELT.

\section{Methodology} \label{sec:methods}
We apply the comprehensive model suite described in \citet{Herbst-etal-2019b} to study the habitability of Prox Cen b as influenced by the strong stellar activity of its host star. 
\subsection{Initial Atmosphere}\label{sec:init}
We use our 1D climate-chemistry model (1D CCM) \citep[see, e.g.,][]{Rauer2011,vonParis2015,Scheucher2018} to calculate initial climate and neutral atmospheric composition for Prox Cen b without SEP or GCR impacts. The stellar energy spectrum is taken from the Virtual Planetary Laboratory Spectral Database\footnote{\url{http://depts.washington.edu/naivpl/content/spectral-databases-and-tools}}, described in \citet{meadows2018}, and the incoming stellar irradiation is scaled to Prox Cen b's distance of 0.0485 AU. To build our planet, we start with the observed minimum mass of 1.27 M$\rm _{E}$ and use the mass-radius relationship from \citet{valencia2007} with an Earth-like ice-mass-fraction of 0.1\%, which results in a radius R of 1.065 R$\rm _{E}$ and a surface gravity g$\rm _{surf}$ of 10.98 ms$^{-2}$ via m/R$^2$.\\
We start with the Earth US standard 1976 atmosphere (Committee on Extension to the Standard Atmosphere~-~COESA) and increase the surface pressure to p$\rm _{surf}$~=~1.119 bar, in order to maintain Earth's atmospheric mass. \rev{We use Earth reference surface fluxes in our model similar to \citet{segura2005} and \citet{meadows2018} which result in the modern Earth 1976 surface mixing ratios of 1.5~ppm~CH$_4$, 190~ppm~CO and 270~ppm~N$_2$O. This is achieved in our chemical scheme with 8.2x10$^{10}$ molec./cm$^2$/s CH$_4$, 1.8x10$^{11}$ molec./cm$^2$/s CO and 1.1x10$^{9}$ molec./cm$^2$/s N$_2$O. These fluxes together with initial surface mixing ratios of 500~ppm~H$_2$, 21.1\%~O$_2$ and 0.934\%~Ar are used as boundary conditions for all our Prox Cen b runs.} With cloud-free conditions and a basaltic surface albedo of A$\rm _{surf}$=0.13, we increase the CO$_2$ amount in the atmosphere \rev{by replacing N$_2$ with CO$_2$ step-by-step. Our aim is to investigate potential surface habitability. We start with 5~\% CO$_2$ (72.4~\% N$_2$) and increase in steps of 5\% up to 20\% CO$_2$ (57.4~\% N$_2$ respectively)}. The tropospheric temperatures and water amounts are calculated via adiabatic lapse rates after \citet{manabe1967} with a surface relative humidity, RH~=~80\%. \rev{For kinetic transport in the chemistry calculations we use eddy diffusion parameterized for Earth after \citet{massie1981}.}
\subsection{Galactic and stellar cosmic ray spectra}
To model the impact of energetic particles on planetary atmospheres both galactic cosmic rays (GCRs) and SEPs have to be considered.\rev{Since Prox Cen is our nearest neighbor it is reasonable to assume the same local interstellar medium (LISM) conditions as for our Sun. However, analytical studies by \citet{Struminsky2017} and \citet{Sadovski2018} showed that GCRs with energies below 1 TeV are not able to reach Prox Cen b. Besides, the flux of such high energetic GCR particles in the LISM is vanishingly small and can be neglected to a first-order approximation. However due to the high stellar activity of Prox Cen, SEPs most likely have a strong impact on Prox Cen b. For example,} \citet{Howard-etal-2018} most recently found strong X-ray flares with intensities up to 6.02 W/m$^2$ on Prox Cen. \citet{Herbst-etal-2019a} suggested that such high flare intensities correspond to stellar proton fluences of $10^8 - 10^{14}$ protons/(cm$^2$ sr s) around Prox Cen at 0.048~AU. To derive our actual particle spectrum, we scale from a well-known measured event spectrum. Our study is based on one of the strongest events measured on Earth, the ground level enhancement (GLE) of February 1956 (GLE05, see upper panel of Fig.~\ref{fig7-2-2}).
\subsection{Planetary magnetic field and atmospheric ionization}
\rev{From Earth we know that low-energetic particles are deflected by the geomagnetic field which acts as an additional particle filter. Thus, the CR flux at the top-of-the-magnetosphere (TOM) is not the same as that at the top-of-the-atmosphere (TOA), which further depends on the magnitude and geometry of the planetary magnetic field \citep[see, e.g.,][]{Herbst-etal-2013}.}

Assuming an Earth-like magnetic field, the so-called cutoff rigidity $R_c$, an equivalent to the energy that particles require to enter the atmosphere at a given location, is computed with PLANETOCOSMICS \citep[see, e.g.,][]{desorgher2006planetocosmics}. \rev{As a first order approximation, in this study we utilize the implemented International Geomagnetic Reference Field \citep[IGRF, see, e.g.,][]{Thebault2015} model to describing the modern-Earth internal terrestrial magnetic field. The globally distributed cutoff rigidity values have been modeled based on the highest planetary disturbance value (k$_p>$7). However, since no information on the spatial resolution of the exoplanetary atmospheric transmission is available, in this study, we assume planetary mid-latitudes around 60$^\circ$, corresponding to a mean cutoff rigidity of 1.11 GV (around 510 MeV), as indicated by the dashed line in the upper panel of Fig.~\ref{fig7-2-2}.} As can be seen, this cutoff separates the TOM from the TOA spectrum.

\rev{However, we note that the magnetospheric structure of Prox Cen b could be different from that of the Earth, for example, due to strong Joule heating caused by fast stellar winds impinging on the upper planetary atmosphere \citep[see, e.g.,][]{Cohen-etal-2014}.}

\rev{Nevertheless, energetic charged particles that reach the planetary atmosphere will lose energy due to collisions with the surrounding atmospheric constituents, resulting in an ionization of the upper planetary atmosphere. Further, interactions with, for example, nitrogen, oxygen, or argon atoms might trigger the development of a secondary particle shower and associated  photochemical  effects. The deeper a particle is able to enter into the atmosphere, the more likely a collision with these species becomes. The generated secondary particles may further interact, resulting in the formation of atmospheric particle cascades \citep[see, e.g.,][]{Dorman-etal-2004} and an altitude-dependent atmospheric ionization. This, however, strongly depends on the type and energy of the primary particle, the atmospheric altitude, and the location.}

\rev{Neglecting GCRs and their astrospheric modulation, the SEP-induced ionization rates $Q$ can numerically be described by}
\begin{equation}
    Q(E_c,x) = \sum_i \int_{E_c}^{E_u} J_i(E) \cdot Y_i(E,x) dE,   
\end{equation}
\rev{with $i$ representing the primary particle type, $E_u$ the upper energy of the stellar particle event, $J_i$ the stellar differential particle event spectrum, and $Y_i$ the so-called atmospheric ionization yield function given as $\alpha\cdot\frac{1}{W_{\mathrm{ion}}}\frac{\Delta E_i}{\Delta x}$, where $\frac{\Delta E_i}{\Delta x}$ reflects the mean specific energy loss, while $W_{\mathrm{ion}}$ represents the atmospheric ionization energy \citep[see, e.g.,][]{Porter-etal-1976, Wedlund-etal-2011}.}

The event-dependent atmospheric ionization rates are modeled with the newly-developed Atmospheric Radiation Simulator \citep[AtRIS, see][]{Banjac-etal-2019a}, utilizing the provided planetary and atmospheric conditions, as well as the particle event spectrum.
\picwide{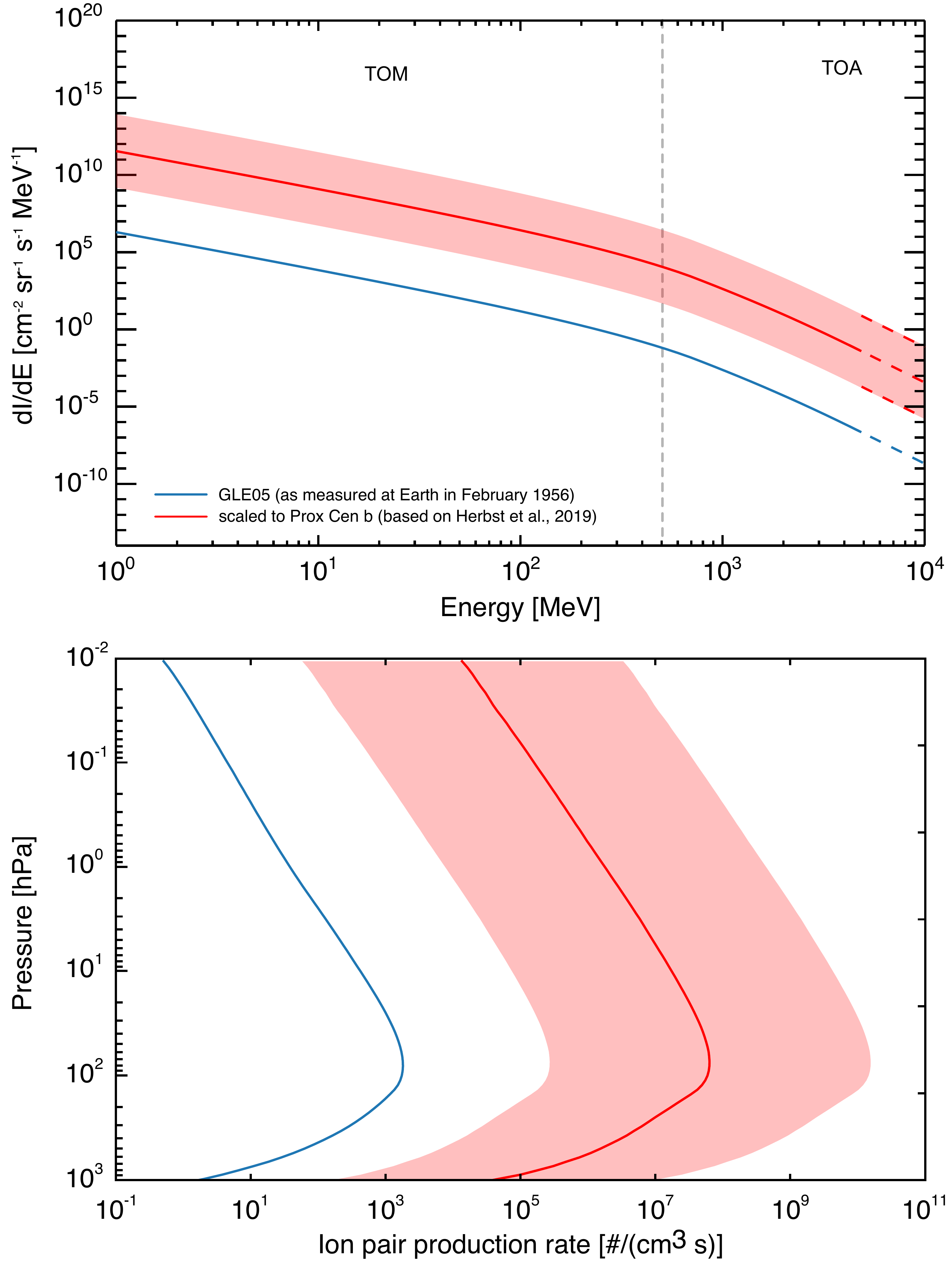}{ Upper panel: Observed GLE event spectrum of GLE05 (blue) compared to the scaled event spectrum at Prox Cen b (red) including the corresponding error band (red shaded region) and cutoff energy on top of the atmosphere (TOA) (grey dashed). Lower panel: Corresponding event-induced atmospheric ion pair production rates at Earth (blue) and at Prox Cen b (red).}
\subsection{Impact on atmospheric ionization, neutral-chemistry, and climate}
The impact of ionization on neutral composition is modeled with the 1D Exoplanetary Terrestrial Ion Chemistry model (ExoTIC) (\cite{Herbst-etal-2019b}, based on the UBIC model described, e.g., in \cite{winkler2009,sinnhuber2012,nieder2014}), taking global averages of the particle-induced ionization calculated by AtRIS \citep[][]{Banjac-etal-2019a}. ExoTIC considers 60 neutral and 120 ion species for neutral, neutral-ion and photochemical reactions. Primary ions as well as excited species are provided from the ionization, dissociation and dissociative ionization of O$_2$, N$_2$ and O; an increase in CO$_2$ mixing ratios may therefore lower the amount of primary ions from N$_2$, O$_2$ and O, but CO$_2$ dissociative ionization is not considered.\\
Both particle-induced ionization and the production and loss rates of NOx (N($^2$D), N($^4$S), NO, NO$_2$, NO$_3$, N$_2$O$_5$), HOx (H, OH), HNO$_3$, H$_2$O, O$_3$, O($^3$P), and O($^1$D) per initial ion, are then transferred from ExoTIC to the 1D CCM, which produces atmospheric climate and composition under SEP and GCR bombardment as input for the next iteration with AtRIS and ExoTIC. Planetary conditions are considered to be in equilibrium when neither ionization, redistribution rates, nor atmospheric conditions change.
\subsection{Spectral characteristics}
For spectral analysis we supply output (p, T, composition) from the coupled model suite as input into the "Generic  Atmospheric  Radiation  Line-by-line  Infra-red  Code" \textsc{(GARLIC)} \citep[e.g.][]{schreier2014, schreier2018agk, schreier2018ace} using HITRAN 2016 \citep{GORDON20173}, CKD continua derived from \cite{clough1989}, visible and near infra-red (IR) cross sections from the Mainz Spectral Atlas \citep{spectralatlas}, and Rayleigh scattering parameterization from \citet{sneep2005},\citet{marcq2011} and \citet{murphy1977}.

\section{Results} \label{sec:results}

\picwide{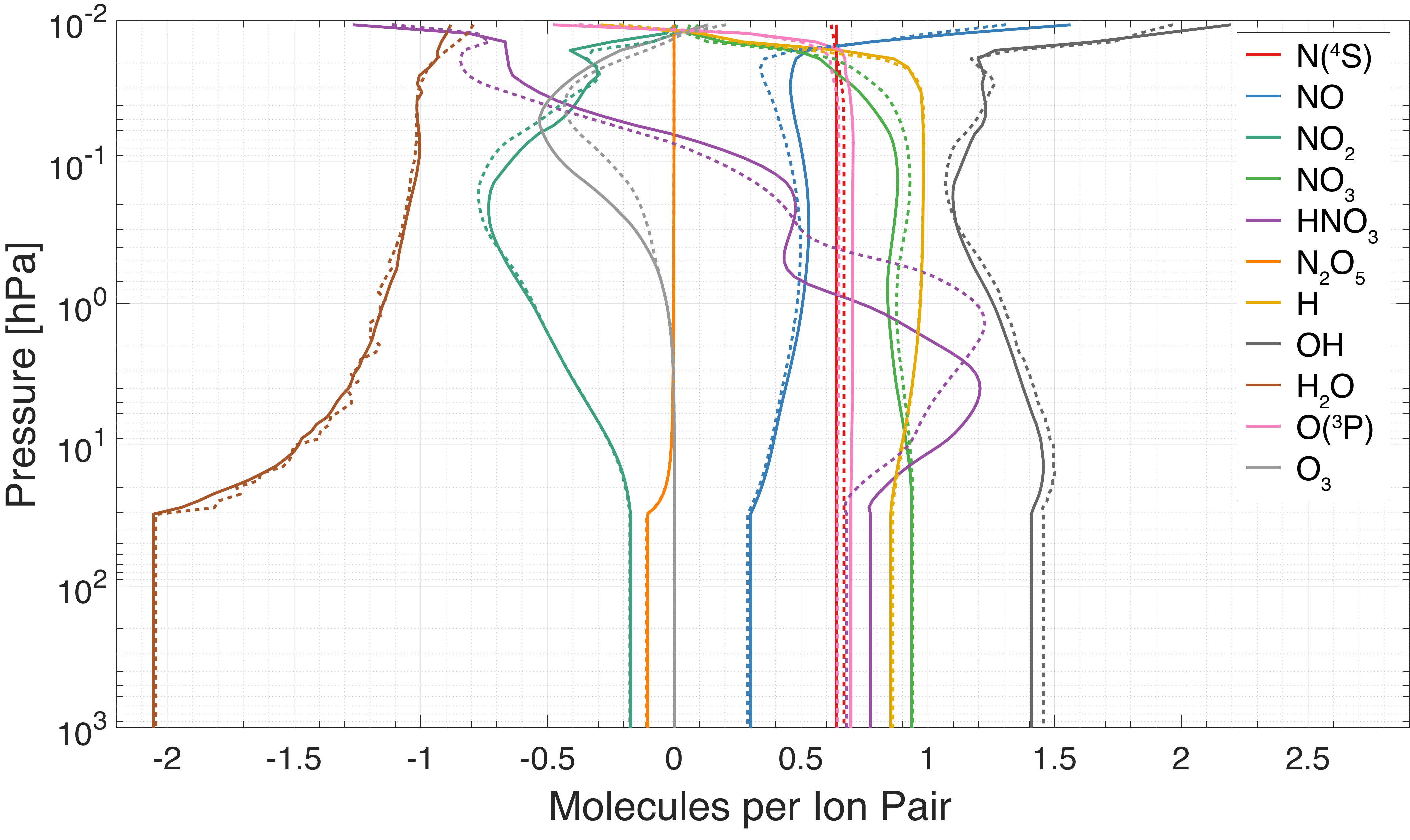}{Production and loss rates of neutral species due to ion-chemistry processes calculated with ExoTIC for the 5\%~CO$_2$ (solid) and 20\%~CO$_2$ (dotted) atmospheres for the GLE05 event scaled to Prox Cen b. Below 30 hPa, rates are set to iso-profiles.}
\picwide{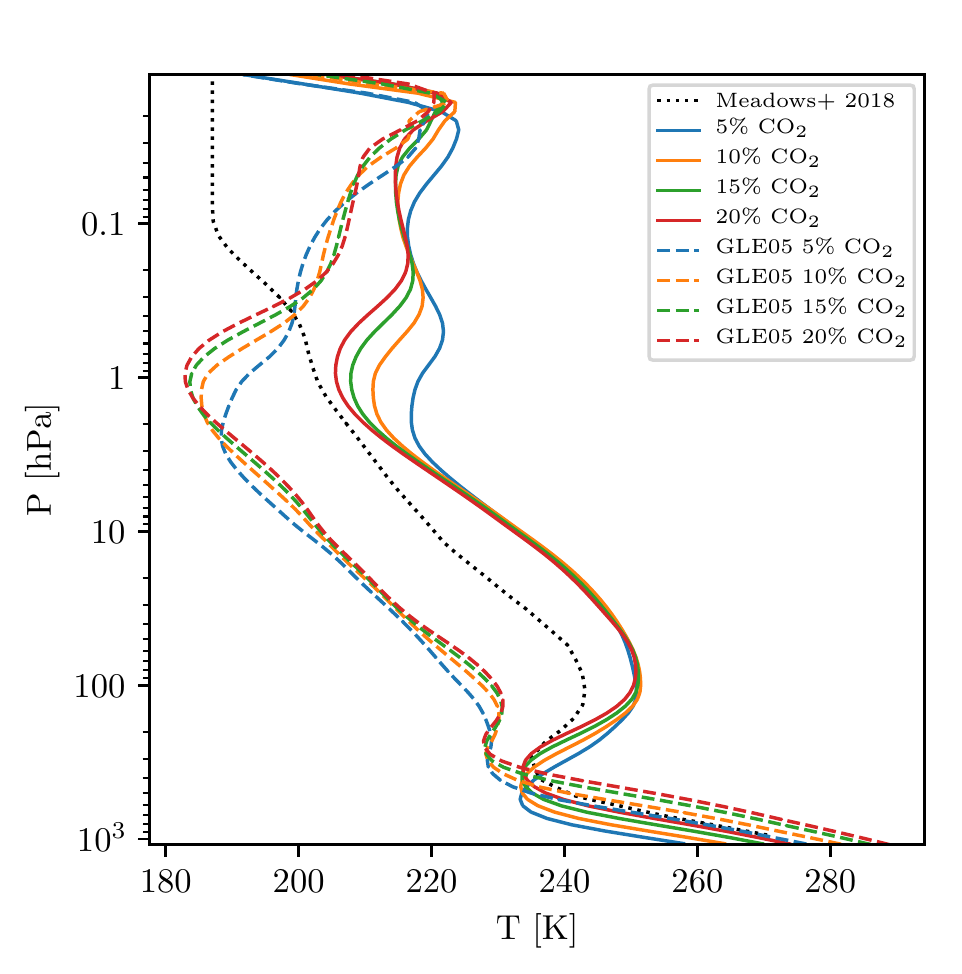}{Atmospheric temperature profiles for an Earth-like Prox Cen b with varying amounts (vmr) of atmospheric CO$_2$ (colors). We compare results for a virtually quiescent Prox Cen (solid) with a high flaring host star (dashed). Overplotted (black dotted) are results from \cite{meadows2018}.}
\picwide{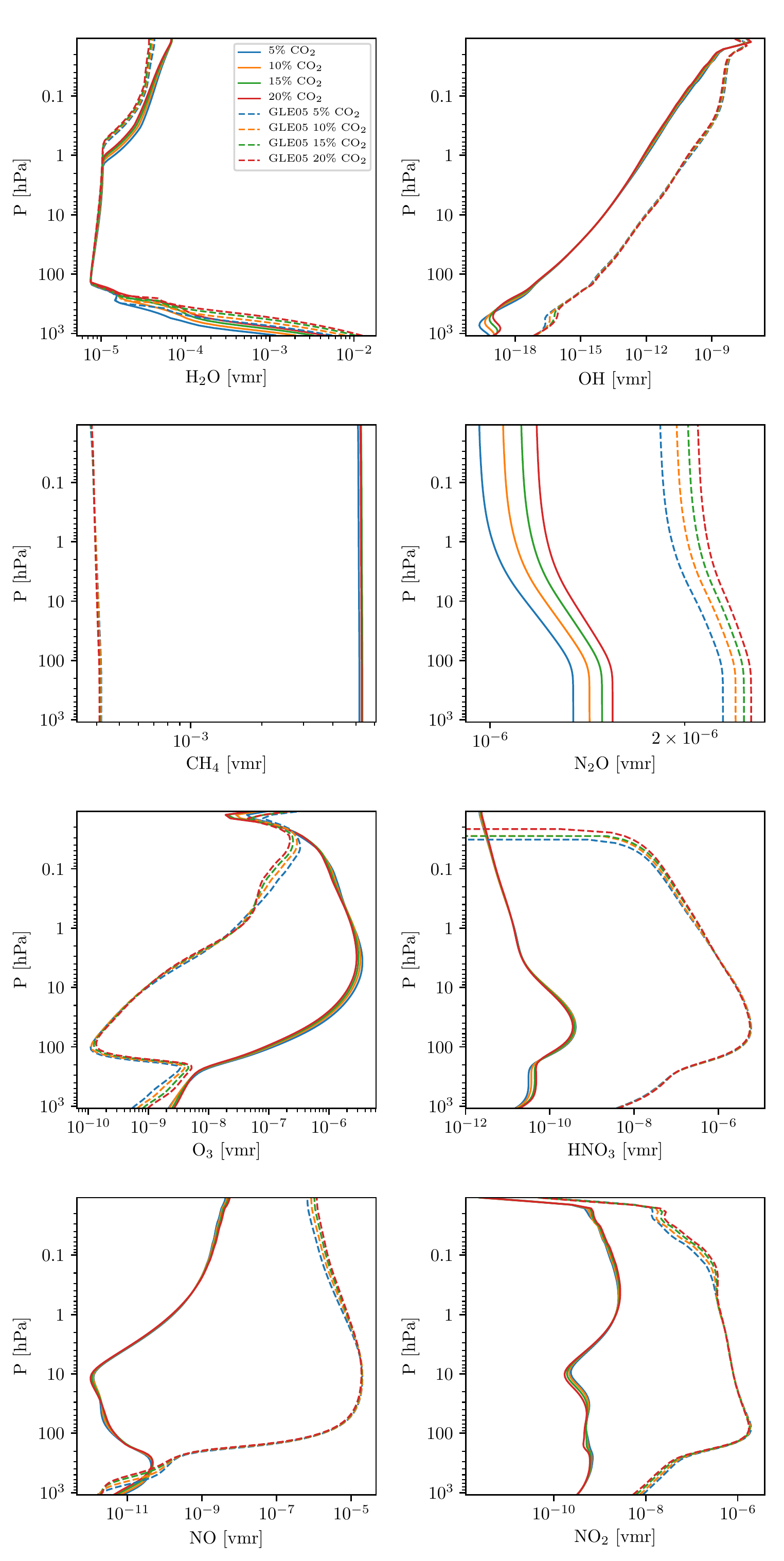}{Volume mixing ratios of \rev{H$_2$O, OH, CH$_4$, N$_2$O, O$_3$, HNO$_3$, NO, and NO$_2$}, for the scenarios of Fig.~\ref{temp}.}
\picwide{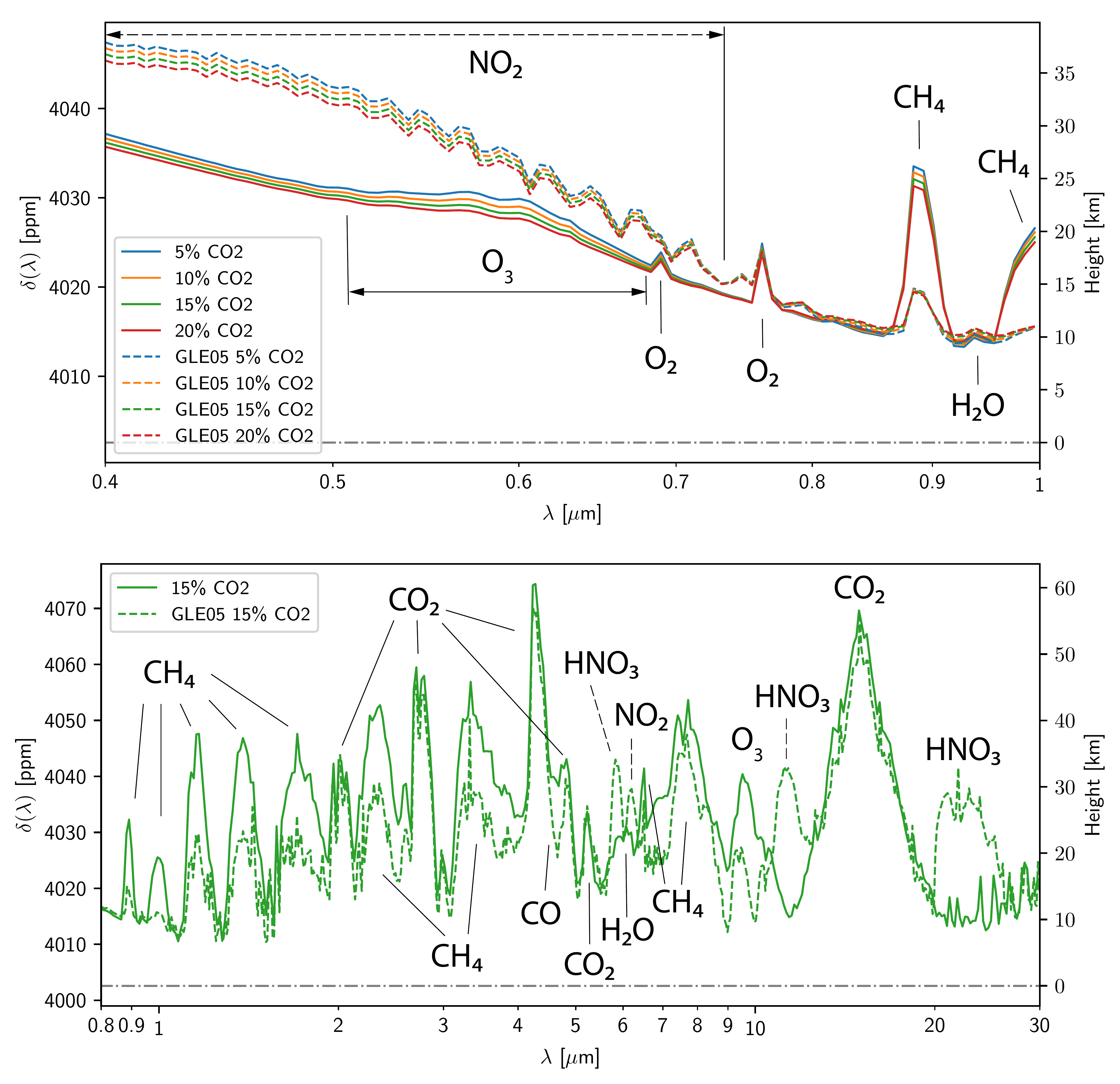}{\underline{Upper:} Transmission spectra (R~=~100) from 400-1000~nm for the scenarios shown in Fig.~\ref{temp} and \ref{cenb-project-paper-comp-vmr__plot}. \underline{Lower:} IR transmission spectral comparison for the two 15\%~CO$_2$ scenarios, the quiescent host star (solid), and GLE05 conditions (dashed).}

The upper panel of Fig.~\ref{fig7-2-2} shows the energy spectrum of GLE05 at Earth (blue) and its scaling to the energy spectrum at Prox Cen b \citep[red, see][]{Herbst-etal-2019a}; the lower panel displays the corresponding cosmic-ray-induced ion-pair production rate calculated for the GLE05 from Earth scaled to Prox Cen b for the initial atmospheres described in Section \ref{sec:init}. We see the typical ion-pair production peak in the lower stratosphere due to increasing atmospheric density. \\
Figure~\ref{kit} shows the ion-chemistry response for the species submitted from the ion-chemistry model to the 1D CCM. Tropospheric values are set to constants based on the lowermost stratospheric values. Rates for H and N($^4$S) are similar to Earth-like values of 1 and 0.6 respectively \citep[compare, e.g.,][]{Herbst-etal-2019b,sinnhuber2012}, while values, e.g., for NO and OH are different. The NO formation rate is smaller than on Earth ($<$0.5 compared to $\sim$0.58), with values for the 20$\%$ CO$_2$ atmosphere being smaller than for the 5$\%$ CO$_2$ case, indicating that this difference is due to the change in bulk atmosphere. The OH formation rate is distinctly larger than for Earth ($\sim$1.5 compared to $\le$1), with values for the  20$\%$ CO$_2$ case larger than for the 5$\%$ CO$_2$ case, again indicating that this difference is due to the change in bulk atmospheric composition. Further analysis shows that the formation rates of NO and OH from positive ion chemistry reactions are similar to Earth values, while those from negative ion chemistry are very different (not shown). Note the strong changes of the formation rates of HNO$_3$ and O$_3$ between the 5$\%$ and 20$\%$ CO$_2$ cases. This might indicate that the different formation rates of OH and NO on Earth may be due to a different composition of negative NO$_3^-$ containing cluster ions which also play a role in HNO$_3$ formation via recombination \citep[e.g.,][]{sinnhuber2012}. This demonstrates the importance of considering the full ion chemistry even for Earth-like (O$_2$-N$_2$) atmospheres.
\\
Figure~\ref{temp} shows temperature profiles for scenarios having Earth-like atmospheres but with varying amounts of CO$_2$ for quiescent stellar conditions (solid lines) and with GCRs and SEPs based on GLE05 (dashed lines). Figure~\ref{temp} suggests CO$_2$ greenhouse warming in the lower atmosphere together with associated mesospheric cooling. Associated with the weak stellar irradiation, results suggest 20\% mixing ratio by volume (vmr) of CO$_2$ is needed in the non-flare cases to achieve global average temperatures that support liquid surface water. In the middle atmosphere, results suggest that the temperature is not sensitive to changes in CO$_2$. A comparison of our 5\%~CO$_2$ run (solid-blue) with \citet{meadows2018} who assumed an Earth-like Prox Cen b with 5\%~CO$_2$ and who did not consider cosmic rays (black-dotted), generally shows reasonable agreement, although some differences arise due to \rev{the different assumptions used for} CH$_4$ \rev{surface fluxes. Our model with post-industrial surface fluxes (solid blue line, 5\%~CO$_2$ in Fig.~\ref{cenb-project-paper-comp-vmr__plot}) yields CH$_4$ concentrations of $\sim$5300 ppm, while \citet{meadows2018} reported $\sim$1000 ppm with lower pre-industrial CH$_4$ fluxes.} In our model this leads to a stronger CH$_4$ anti-greenhouse, hence lower surface temperatures with a warmer stratosphere.\\
Figure~\ref{temp} additionally shows (dashed lines) atmospheric temperatures for flaring scenarios of Prox Cen with the calculated GLE05 induced ionization from Figure~\ref{fig7-2-2} and chemical production/loss rates from Figure~\ref{kit}. Surface CH$_4$ here is decreased from 5300 down to 400 ppm (although CH$_4$ vmr are still essentially iso-profiles \rev{as shown in Fig.~\ref{cenb-project-paper-comp-vmr__plot}}), yielding a weaker anti-greenhouse effect, resulting in higher surface temperatures for the flaring cases. Interestingly, all four tested CO$_2$ concentrations are sufficient for the flaring cases to warm the surface and
lead to habitable conditions. The 15\%~CO$_2$ scenario outputs Earth-like 288~K global average surface temperatures. \\
Figure~\ref{cenb-project-paper-comp-vmr__plot} shows atmospheric vmr profiles of \rev{H$_2$O, OH, CH$_4$, N$_2$O, O$_3$, HNO$_3$, NO, and NO$_2$} for the quiescent and flaring cases from Figure~\ref{temp}. \rev{Our CH$_4$ abundances are strongly increased compared to Earth. This arises \citep[see ][]{segura2005} due to lower UVB radiation which lowers O$_3$ photolysis; this then lowers O($^1$D) (a product of O$_3$ photolysis), which lowers the rate of H$_2$O+O($^1$D), hence lowers OH and increases CH$_4$. For the flaring cases, however, CH$_4$ is effectively reduced by high amounts of cosmic-ray-induced OH, via $\rm CH_4 + OH \longrightarrow CH_3 + H_2O$.} Stratospheric ozone strongly decreases similarly to \citet{Scheucher2018}. Increases in OH and NO (which can remove stratospheric ozone and stimulate smog ozone) for the flaring cases are a result of the cosmic-ray-induced production rates from ion redistribution into neutral species. HNO$_3$ increases by up to four orders of magnitude compared to the non-flaring case, mostly due to photochemistry (its main in-situ source is via the reaction: $\rm NO_2 + OH + M \longrightarrow HNO_3 + M$ ['M' refers to any third body]. The gas-phase precursors of HNO$_3$ i.e. NO$_2$ and OH, are stimulated by cosmic-ray-induced chemistry. H$_2$O does not show significant changes in molecular abundance. \rev{NO$_2$ shows up to four orders of magnitude increased concentrations for the flaring compared to quiescent case. Fig.~\ref{kit} shows cosmic-ray-induced ion-chemical destruction of NO$_2$. NO$_2$ is incorporated into cluster ions which release other N-containing species under recombination, either NO, NO$_3$ or HNO$_3$, which then form NO$_2$ in a multitude of secondary neutral gas-phase reactions, overwhelming the apparent loss. One important source is the higher UVB radiation environment in the lower-middle atmosphere because of the lowered O$_3$ concentration, hence reduced UVB absorption. This increases photolysis of, most importantly, HNO$_3$, HO$_2$NO$_2$ and N$_2$O$_5$, producing large amounts of NO$_2$. Further, NO$_2$ is a direct product of the O$_3$ destruction mechanism starting with $\rm NO + O_3 \longrightarrow NO_2 + O_2$, and cosmic-ray-induced NOx and HOx also contribute significantly via the two reactions $\rm NO + HO_2 \longrightarrow NO_2 + OH$ and $\rm NO + NO_3 \longrightarrow 2 NO_2$. N$_2$O, on the contrary, does not show changes due to cosmic rays significant enough to show up in our simulated spectra (Fig.~\ref{spectra}). The $\sim$65\% increase in overall abundance in Fig.~\ref{cenb-project-paper-comp-vmr__plot} for our flaring cases compared to the quiescent runs is a little counter intuitive at first, because with the large decrease in O$_3$ amounts the overall UVB and UVC radiation environment, i.e. photolysis of other species, increases. Our anaylsis showed that UVC fluxes $<$198~nm is decreased significantly, reducing N$_2$O photolysis - the major N$_2$O sink in our model. From a detailed investigation of changes in molecular abundances of other major absorbers overlapping in wavelength range with N$_2$O photolysis (significant $\sim$175-240~nm), together with their photolysis cross-sections, showed a steep increase of HNO$_3$ and NO$_2$ (and some increase for HO$_2$NO$_2$ below $\sim$20~hPa) photolysis rates, hence shielding of N$_2$O. Small changes in N$_2$O with the variation of CO$_2$ contents are most importantly related to temperature changes, i.e. H$_2$O steam amounts and photolysis rates in the atmosphere.}\\
Figure~\ref{spectra} shows synthetic transmission spectra calculated by GARLIC. The visible to near IR (Fig.~\ref{spectra}, upper panel) is a key region, e.g. for biosignature studies of, for example, O$_2$ with the ELT \citep[e.g.][]{snellen2014,rodler2014}. \cite{betremieux2013} noted the importance of including, for example, O$_2$ and O$_3$ absorption in this region, which is overlooked by many biosignature studies, but which is included in our work. 
Our results (Fig.~\ref{spectra}, upper panel) suggest a significant difference between the flaring and quiescent runs - but smaller differences due to changing composition for the individual flaring or quiescent runs. All quiescent runs show the O$_3$ Chappuis bands around 600~nm known from Earth, with minor differences in strength, due to the slightly different strato-/mesospheric O$_3$ amounts. This absorption feature is rather small compared with Earth's atmosphere because of the generally lower O$_3$ amounts of $\sim$~70 Dobson units (DU) for the quiescent case. In the flaring runs, there is a striking broad absorption feature from $\sim$~400~-~700~nm attributed to NO$_2$. Due to high amounts of cosmic-ray-induced NO$_2$, these features act almost as a continuum in the visible (making the sky brownish-red), stronger than the 600~nm O$_3$ feature previously mentioned, and could potentially be misinterpreted as a steeper Rayleigh slope. We propose the NO$_2$ feature as a spectral "marker" of N$_2$-O$_2$ atmospheres which are subject to cosmic rays \citep[see also][]{Airapetian2017b}. \\
The IR spectrum (Fig.~\ref{spectra}, lower panel) shows a direct comparison of the 15\%~CO$_2$ runs, where the flaring case leads to Earth-like temperatures of 288~K in Fig.~\ref{temp}. Again, the effect of varying composition was minor (not shown). As expected, the CH$_4$ features in the mid IR, coinciding with H$_2$O features, are significantly reduced because of reduced CH$_4$ in the flaring runs. Similar to \citet{tabataba2016} and \citet{Scheucher2018}, strong flare features from HNO$_3$ occur around 11~$\mu$m and 21~$\mu$m, but the O$_3$ absorption feature around 9.6~$\mu$m is greatly reduced. \\
Around 5.3~$\mu$m, we see a narrow but distinct absorption feature around 10~ppm (in $\delta(\lambda)$) above the lower-atmosphere H$_2$O absorption background. On modern Earth, there is a band at 5.3~$\mu$m from NO via stratospheric N$_2$O oxidation which varies considerably with geomagnetic activity, due to fast solar wind, CIRs and CMEs, as well as from extra NO production in the upper mesosphere / lower thermosphere from far-UV and EUV photoionization, and particle impact ionization. Therefore the strength of this NO feature can be treated as an SEP indicator in Earth's atmosphere \citep[see, e.g.,][]{Airapetian2017b}. However, in our modeled spectra, this feature (peaking at 5.35~$\mu$m) has the same strength for both the flaring and the quiescent cases. We identified this as a weak absorption band of CO$_2$, which generally receives little attention in the literature e.g. because of the much stronger CO$_2$ bands in the spectral vicinity at 4.3 and 4.8~$\mu$m. This 5.3~$\mu$m spectral feature could therefore lead to misinterpretation of future observations because of the above discussed clear correlation with flares in Earth's atmosphere.\\
It is striking that there is no single significant H$_2$O feature in the spectrum. We tested this up to JWST-like spectral resolutions of R~=~3000 in the near IR (not shown) by removing H$_2$O absorption from the spectrum. The resulting spectra showed differences of no more than 2~ppm (in $\delta(\lambda)$) in selected CH$_4$ window regions. H$_2$O concentrations in an Earth-like troposphere decrease rapidly with height \rev{with the existence of a cold trap}, hence major H$_2$O absorption features arise lower down in the atmosphere. For example, in the 15~$\mu$m CO$_2$ band H$_2$O would become optically thick at 28 km in our runs shown in Fig.~\ref{spectra}, whereas CO$_2$ makes the atmosphere opaque at heights of 61 km in transmission; thus, the absorption due to H$_2$O could not be measured. Similarly, between 1-4~$\mu$m all H$_2$O absorption bands are overlapped by strong CH$_4$ absorption which makes the atmosphere opaque at around 40~km. \rev{It is noteworthy, that in the absence of SEPs there is a weak feature from the H$_2$O absorption band around 6~$\mu$m as pointed out by \citet{meadows2018} and indicated in Fig~\ref{spectra}, which would be difficult to distinguish from background noise in our simulations.}
\section{Summary} \label{sec:sum}
We performed a habitability study of Prox Cen b assuming an Earth-like atmosphere, focusing on the influence of SEPs from stellar flares upon spectral transmission features. We applied our extensive model suite discussed in \citet{Herbst-etal-2019b}, which includes: the calculation of SEP fluxes and spectral energy distributions; their precipitation through a planetary magnetosphere and atmosphere; cosmic-ray-induced atmospheric ionization; ion redistribution within the atmosphere; and atmospheric climate and neutral composition. Our results suggest a few words of caution for the analysis of spectra from future observations of such planets around M-dwarf stars. \\
Given the general lack of in-situ particle measurements from flaring events of Prox Cen, we scale the known SEP flux of GLE05 measured at Earth to the intensity of the measured flare \citep[][]{Herbst-etal-2019a} and the orbit of Prox Cen b, then calculate the ion-pair production in a theoretical Earth-like atmosphere with elevated CO$_2$ amounts in order to warm the planet at its position of 0.65 TSI. Together with the production and loss rates of neutral species from the redistribution of cosmic-ray-induced ions, we find steady-state climate and atmospheric compositions that enable liquid surface water. Our results suggest that SEPs might play a crucial role in efficiently warming such planets, which would otherwise be too cold to be habitable (by reducing the anti-greenhouse gas CH$_4$). Numerous studies \citep[e.g.][]{segura2005,Rauer2011} have suggested that planets with Earth-like biomasses could develop methane abundances orders of magnitude higher than on Earth because of the different stellar spectrum \rev{(especially the reduced UVB),} which can then lead to anti-greenhouse methane cooling (see Fig.~\ref{temp}).\\
By including SEPs from stellar flares, strong production of cosmic-ray-induced OH efficiently reduces CH$_4$, hence results in temperate global average surface temperatures for all our test atmospheres. This result is exciting because strong SEPs and flares are generally considered to be disfavorable for planetary habitability, because of potentially increased atmospheric escape and increases in surface dosage. We suggest that the habitability of Earth-like planets near the outer HZ of cooler stars \rev{could depend strongly on both, CH$_4$ and CO$_2$}. Methane, in turn, can be sensitive to cosmic-ray-induced chemistry.\\

We identify some key features in our synthetic planetary transmission spectra (Fig.~\ref{spectra}), some related to SEPs, and note some cautions related to the possible misinterpretation of future spectral observations.\\

$\bullet$ N$_2$-O$_2$-dominated atmospheres could produce large cosmic-ray-induced amounts of NO$_2$, strongly absorbing between 400 - 700~nm. This could lead to erroneous Rayleigh slope extraction from observations, which could potentially influence estimates of e.g. bulk atmospheric composition or/and atmospheric dust and aerosols / hazes. \\

$\bullet$ No significant H$_2$O absorption feature can be seen in our modeled transmission spectra. This is because H$_2$O absorption bands are strongly overlapped by CH$_4$ and CO$_2$ absorption, which occurs higher up in the atmosphere, thus making these spectral regions opaque for radiation. One would need high-resolution spectroscopy to infer H$_2$O from near-mid IR transmission spectroscopy of such planets. \\

$\bullet$ Results suggest HNO$_3$ spectral features become apparent in the mid and far IR, similar to results discussed in \citet{tabataba2016,Scheucher2018}. These features might not only be an indicator for SPEs bombarding a planet with an N$_2$-O$_2$ atmosphere, but could potentially be used as an indirect hint for the presence of H$_2$O where one cannot see its absorption features in the spectrum.\\

$\bullet$ O$_3$ features are rather weak for scenarios around quiescent mid-late type M-dwarfs. This arises mainly due to photochemical responses due to the different stellar spectrum \cite[see e.g.][]{segura2005,grenfell2012}. The weakened ozone features are then even further reduced by SEPs. \\

$\bullet$ We calculate a 5.3~$\mu$m feature with similar strength in both our non-flaring as well as flaring star runs. This comes from a weak 5.35~$\mu$m absorption band of CO$_2$ and overlaps strongly with the 5.3~$\mu$m NO absorption feature that is evident after SEP events on Earth. Therefore one might misinterpret the presence of such a feature as an Earth-like N$_2$-O$_2$ atmosphere under strong SEP bombardment.\\

In a nutshell, we applied our extensive model suite to study the influence of cosmic rays on Prox Cen b as a potentially Earth-like planet and identified some interesting spectral features with strong potential for general characterization and misinterpretation of atmospheric compositions of Earth-like planets in the HZ around M-stars.

\acknowledgments
MSch acknowledges support from DFG project RA 714/9-1. KH and JLG acknowledge the International Space Science Institute and the supported International Team 464: \textit{The  Role  Of  Solar  And  Stellar  Energetic  Particles  On (Exo)Planetary Habitability (ETERNAL, \url{http://www.issibern.ch/teams/exoeternal/})}. VS and MSi acknowledge support from DFG project SI 1088/4-1. FS acknowledges support from DFG project SCHR 1125/3-1.

\software{GARLIC \citep{schreier2014,schreier2018agk,schreier2018ace}, HITRAN2016 \citep{GORDON20173}}, MPI Mainz Spectral Atlas \citep{spectralatlas}

\bibliographystyle{aasjournal}

\begin{thebibliography}{}
\expandafter\ifx\csname natexlab\endcsname\relax\def\natexlab#1{#1}\fi
\providecommand{\url}[1]{\href{#1}{#1}}

\bibitem[{Airapetian {et~al.}(2017{\natexlab{a}})Airapetian, Glocer, Khazanov,
  Loyd, France, Sojka, Danchi, \& Liemohn}]{Airapetian2017}
Airapetian, V.~S., Glocer, A., Khazanov, G.~V., {et~al.} 2017{\natexlab{a}},
  The Astrophysical Journal, 836, L3.
\newblock \url{https://doi.org/10.3847%2F2041-8213%2F836%2F1%2Fl3}

\bibitem[{Airapetian {et~al.}(2017{\natexlab{b}})Airapetian, Jackman, Mlynczak,
  Danchi, \& Hunt}]{Airapetian2017b}
Airapetian, V.~S., Jackman, C.~H., Mlynczak, M., Danchi, W., \& Hunt, L.
  2017{\natexlab{b}}, Scientific Reports, 7, 14141.
\newblock \url{https://doi.org/10.1038/s41598-017-14192-4}

\bibitem[{Anglada-Escud{\'e} {et~al.}(2016)Anglada-Escud{\'e}, Amado, Barnes,
  Berdi{\~n}as, Butler, Coleman, de~la Cueva, Dreizler, Endl, Giesers, Jeffers,
  Jenkins, Jones, Kiraga, K{\"u}rster, L{\'o}pez-Gonz{\'a}lez, Marvin, Morales,
  Morin, Nelson, Ortiz, Ofir, Paardekooper, Reiners, Rodr{\'i}guez,
  Rodriguez-L{\'o}pez, Sarmiento, Strachan, Tsapras, Tuomi, \&
  Zechmeister}]{Anglada2016}
Anglada-Escud{\'e}, G., Amado, P.~J., Barnes, J., {et~al.} 2016, Nature, 536,
  437.
\newblock \url{https://doi.org/10.1038/nature19106}

\bibitem[{Banjac {et~al.}(2019)Banjac, Herbst, \& Heber}]{Banjac-etal-2019a}
Banjac, S., Herbst, K., \& Heber, B. 2019, Journal of Geophysical Research:
  Space Physics, 124, 50.
\newblock \url{https://doi.org/10.1029/2018JA026042}

\bibitem[{Berdyugina \& Kuhn(2019)}]{Berdyugina2019}
Berdyugina, S.~V., \& Kuhn, J.~R. 2019, The Astronomical Journal, 158, 246.
\newblock \url{https://doi.org/10.3847%2F1538-3881%2Fab2df3}

\bibitem[{B{\'{e}}tr{\'{e}}mieux \& Kaltenegger(2013)}]{betremieux2013}
B{\'{e}}tr{\'{e}}mieux, Y., \& Kaltenegger, L. 2013, The Astrophysical Journal,
  772, L31.
\newblock \url{https://doi.org/10.1088%2F2041-8205%2F772%2F2%2Fl31}

\bibitem[{Clough {et~al.}(1989)Clough, Kneizys, \& Davies}]{clough1989}
Clough, S., Kneizys, F., \& Davies, R. 1989, Atmospheric Research, 23, 229 .
\newblock
  \url{http://www.sciencedirect.com/science/article/pii/0169809589900203}

\bibitem[{{Cohen} {et~al.}(2014){Cohen}, {Drake}, {Glocer}, {Garraffo},
  {Poppenhaeger}, {Bell}, {Ridley}, \& {Gombosi}}]{Cohen-etal-2014}
{Cohen}, O., {Drake}, J.~J., {Glocer}, A., {et~al.} 2014, \apj, 790, 57

\bibitem[{Desorgher {et~al.}(2006)Desorgher, Fl{\"u}ckiger, \&
  Gurtner}]{desorgher2006planetocosmics}
Desorgher, L., Fl{\"u}ckiger, E.~O., \& Gurtner, M. 2006, in 36th COSPAR
  Scientific Assembly, Vol.~36

\bibitem[{Dong {et~al.}(2017)Dong, Lingam, Ma, \& Cohen}]{Dong2017}
Dong, C., Lingam, M., Ma, Y., \& Cohen, O. 2017, The Astrophysical Journal,
  837, L26.
\newblock \url{https://doi.org/10.3847%2F2041-8213%2Faa6438}

\bibitem[{{Dorman} {et~al.}(2004){Dorman}, {Pustil'Nik}, {Sternlieb},
  {Zukerman}, {Belov}, {Eroshenko}, {Yanke}, {Mavromichalaki}, {Sarlanis},
  {Souvatzoglou}, {Tatsis}, {Iucci}, {Villoresi}, {Fedorov}, {Shakhov}, \&
  {Murat}}]{Dorman-etal-2004}
{Dorman}, L.~I., {Pustil'Nik}, L.~A., {Sternlieb}, A., {et~al.} 2004, IEEE
  Transactions on Plasma Science, 32, 1478

\bibitem[{Gordon {et~al.}(2017)Gordon, Rothman, Hill, Kochanov, Tan, Bernath,
  Birk, Boudon, Campargue, Chance, Drouin, Flaud, Gamache, Hodges, Jacquemart,
  Perevalov, Perrin, Shine, Smith, Tennyson, Toon, Tran, Tyuterev, Barbe,
  Császár, Devi, Furtenbacher, Harrison, Hartmann, Jolly, Johnson, Karman,
  Kleiner, Kyuberis, Loos, Lyulin, Massie, Mikhailenko, Moazzen-Ahmadi,
  Müller, Naumenko, Nikitin, Polyansky, Rey, Rotger, Sharpe, Sung, Starikova,
  Tashkun, Auwera, Wagner, Wilzewski, Wcisło, Yu, \& Zak}]{GORDON20173}
Gordon, I., Rothman, L., Hill, C., {et~al.} 2017, Journal of Quantitative
  Spectroscopy and Radiative Transfer, 203, 3 , hITRAN2016 Special Issue.
\newblock
  \url{http://www.sciencedirect.com/science/article/pii/S0022407317301073}

\bibitem[{{Grenfell} {et~al.}(2012){Grenfell}, {Grie{\ss}meier}, {von Paris},
  {Patzer}, {Lammer}, {Stracke}, {Gebauer}, {Schreier}, \&
  {Rauer}}]{grenfell2012}
{Grenfell}, J.~L., {Grie{\ss}meier}, J.-M., {von Paris}, P., {et~al.} 2012,
  Astrobiology, 12, 1109

\bibitem[{{Herbst} {et~al.}(2013){Herbst}, {Kopp}, \&
  {Heber}}]{Herbst-etal-2013}
{Herbst}, K., {Kopp}, A., \& {Heber}, B. 2013, Annales Geophysicae, 31, 1637

\bibitem[{{Herbst} {et~al.}(2019{\natexlab{a}}){Herbst}, Papaioannou, Banjac,
  \& {Heber}}]{Herbst-etal-2019a}
{Herbst}, K., Papaioannou, A., Banjac, S., \& {Heber}, B. 2019{\natexlab{a}},
  Astron. Astrophys., 621.
\newblock \url{https://doi.org/10.1051/0004-6361/201832789}

\bibitem[{{Herbst} {et~al.}(2019{\natexlab{b}}){Herbst}, {Grenfell},
  {Sinnhuber}, {Rauer}, {Heber}, {Banjac}, {Scheucher}, {Schmidt}, {Gebauer},
  {Lehmann}, \& {Schreier}}]{Herbst-etal-2019b}
{Herbst}, K., {Grenfell}, J., {Sinnhuber}, M., {et~al.} 2019{\natexlab{b}},
  Astron. Astrophys., 631.
\newblock \url{https://doi.org/10.1051/0004-6361/201935888}

\bibitem[{{Howard} {et~al.}(2018){Howard}, {Tilley}, {Corbett}, {Youngblood},
  {Loyd}, {Ratzloff}, {Law}, {Fors}, {del Ser}, {Shkolnik}, {Ziegler}, {Goeke},
  {Pietraallo}, \& {Haislip}}]{Howard-etal-2018}
{Howard}, W.~S., {Tilley}, M.~A., {Corbett}, H., {et~al.} 2018, ArXiv e-prints,
  arXiv:1804.02001

\bibitem[{Keller-Rudek {et~al.}(2013)Keller-Rudek, Moortgat, Sander, \&
  S\"orensen}]{spectralatlas}
Keller-Rudek, H., Moortgat, G.~K., Sander, R., \& S\"orensen, R. 2013, Earth
  System Science Data, 5, 365.
\newblock \url{https://www.earth-syst-sci-data.net/5/365/2013/}

\bibitem[{Lammer {et~al.}(2010)Lammer, Selsis, Chassefière, Breuer,
  Grießmeier, Kulikov, Erkaev, Khodachenko, Biernat, Leblanc, Kallio, Lundin,
  Westall, Bauer, Beichman, Danchi, Eiroa, Fridlund, Gröller, Hanslmeier,
  Hausleitner, Henning, Herbst, Kaltenegger, Léger, Leitzinger, Lichtenegger,
  Liseau, Lunine, Motschmann, Odert, Paresce, Parnell, Penny, Quirrenbach,
  Rauer, Röttgering, Schneider, Spohn, Stadelmann, Stangl, Stam, Tinetti, \&
  White}]{lammer2009}
Lammer, H., Selsis, F., Chassefière, E., {et~al.} 2010, Astrobiology, 10, 45,
  pMID: 20307182.
\newblock \url{https://doi.org/10.1089/ast.2009.0368}

\bibitem[{{Manabe} \& {Wetherald}(1967)}]{manabe1967}
{Manabe}, S., \& {Wetherald}, R.~T. 1967, Journal of Atmospheric Sciences, 24,
  241

\bibitem[{Marcq {et~al.}(2011)Marcq, Belyaev, Montmessin, Fedorova, Bertaux,
  Vandaele, \& Neefs}]{marcq2011}
Marcq, E., Belyaev, D., Montmessin, F., {et~al.} 2011, Icarus, 211, 58 .
\newblock
  \url{http://www.sciencedirect.com/science/article/pii/S0019103510003271}

\bibitem[{Massie \& Hunten(1981)}]{massie1981}
Massie, S., \& Hunten, D. 1981, Journal of Geophysical Research: Oceans, 86,
  9859

\bibitem[{Meadows {et~al.}(2018)Meadows, Arney, Schwieterman, Lustig-Yaeger,
  Lincowski, Robinson, Domagal-Goldman, Deitrick, Barnes, Fleming, Luger,
  Driscoll, Quinn, \& Crisp}]{meadows2018}
Meadows, V.~S., Arney, G.~N., Schwieterman, E.~W., {et~al.} 2018, Astrobiology,
  18, 133, pMID: 29431479.
\newblock \url{https://doi.org/10.1089/ast.2016.1589}

\bibitem[{Murphy(1977)}]{murphy1977}
Murphy, W.~F. 1977, The Journal of Chemical Physics, 67, 5877.
\newblock \url{https://doi.org/10.1063/1.434794}

\bibitem[{{Nieder} {et~al.}(2014){Nieder}, {Winkler}, {Marsh}, \&
  {Sinnhuber}}]{nieder2014}
{Nieder}, H., {Winkler}, H., {Marsh}, D.~R., \& {Sinnhuber}, M. 2014, Journal
  of Geophysical Research (Space Physics), 119, 2137

\bibitem[{{Porter} {et~al.}(1976){Porter}, {Jackman}, \&
  {Green}}]{Porter-etal-1976}
{Porter}, H.~S., {Jackman}, C.~H., \& {Green}, A.~E.~S. 1976, \jcp, 65, 154

\bibitem[{{Rauer} {et~al.}(2011){Rauer}, {Gebauer}, {Paris}, {Cabrera},
  {Godolt}, {Grenfell}, {Belu}, {Selsis}, {Hedelt}, \& {Schreier}}]{Rauer2011}
{Rauer}, H., {Gebauer}, S., {Paris}, P.~V., {et~al.} 2011, \aap, 529, A8

\bibitem[{Reid \& Hawley(2005)}]{Reid2005}
Reid, I.~N., \& Hawley, S.~L. 2005, New Light on Dark Stars: Red Dwarfs,
  Low-Mass Stars, Brown Dwarfs (Springer Berlin Heidelberg),
  doi:10.1007/3-540-27610-6.
\newblock \url{https://doi.org/10.1007/3-540-27610-6}

\bibitem[{Rodler \& L{\'{o}}pez-Morales(2014)}]{rodler2014}
Rodler, F., \& L{\'{o}}pez-Morales, M. 2014, The Astrophysical Journal, 781,
  54.
\newblock \url{https://doi.org/10.1088%2F0004-637x%2F781%2F1%2F54}

\bibitem[{Sadovski {et~al.}(2018)Sadovski, Struminsky, \& Belov}]{Sadovski2018}
Sadovski, A.~M., Struminsky, A.~B., \& Belov, A. 2018, Astronomy Letters, 44,
  324.
\newblock \url{https://doi.org/10.1134/S1063773718040072}

\bibitem[{Scalo {et~al.}(2007)Scalo, Kaltenegger, Segura, Fridlund, Ribas,
  Kulikov, Grenfell, Rauer, Odert, Leitzinger, Selsis, Khodachenko, Eiroa,
  Kasting, \& Lammer}]{scalo2007}
Scalo, J., Kaltenegger, L., Segura, A., {et~al.} 2007, Astrobiology, 7, 85,
  pMID: 17407405.
\newblock \url{https://doi.org/10.1089/ast.2006.0125}

\bibitem[{Scheucher {et~al.}(2018)Scheucher, Grenfell, Wunderlich, Godolt,
  Schreier, \& Rauer}]{Scheucher2018}
Scheucher, M., Grenfell, J., Wunderlich, F., {et~al.} 2018, ApJ, 863, 6.
\newblock \url{https://doi.org/10.3847/1538-4357/aacf03}

\bibitem[{Schreier {et~al.}(2014)Schreier, {Gimeno Garc{\'\i}a}, Hedelt, Hess,
  Mendrok, Vasquez, \& Xu}]{schreier2014}
Schreier, F., {Gimeno Garc{\'\i}a}, S., Hedelt, P., {et~al.} 2014, JQSRT, 137,
  29.
\newblock \url{https://doi.org/10.1016/j.jqsrt.2013.11.018}

\bibitem[{Schreier {et~al.}(2018{\natexlab{a}})Schreier, Milz, Buehler, \& von
  Clarmann}]{schreier2018agk}
Schreier, F., Milz, M., Buehler, S.~A., \& von Clarmann, T. 2018{\natexlab{a}},
  JQSRT, 211, 64.
\newblock \url{https://doi.org/10.1016/j.jqsrt.2018.02.032}

\bibitem[{Schreier {et~al.}(2018{\natexlab{b}})Schreier, St\"adt, Hedelt, \&
  Godolt}]{schreier2018ace}
Schreier, F., St\"adt, S., Hedelt, P., \& Godolt, M. 2018{\natexlab{b}},
  Molec.\ Astrophysics, 11, 1.
\newblock \url{https://doi.org/10.1016/j.molap.2018.02.001}

\bibitem[{Segura {et~al.}(2005)Segura, Kasting, Meadows, Cohen, Scalo, Crisp,
  Butler, \& Tinetti}]{segura2005}
Segura, A., Kasting, J.~F., Meadows, V., {et~al.} 2005, Astrobiology, 5, 706.
\newblock \url{http://adsabs.harvard.edu/abs/2005AsBio...5..706S}

\bibitem[{{Segura} {et~al.}(2010){Segura}, {Walkowicz}, {Meadows}, {Kasting},
  \& {Hawley}}]{segura2010}
{Segura}, A., {Walkowicz}, L.~M., {Meadows}, V., {Kasting}, J., \& {Hawley}, S.
  2010, Astrobiology, 10, 751

\bibitem[{Simon~Wedlund {et~al.}(2011)Simon~Wedlund, Gronoff, Lilensten,
  M\'enager, \& Barth\'elemy}]{Wedlund-etal-2011}
Simon~Wedlund, C., Gronoff, G., Lilensten, J., M\'enager, H., \& Barth\'elemy,
  M. 2011, Annales Geophysicae, 29, 187

\bibitem[{{Sinnhuber} {et~al.}(2012){Sinnhuber}, {Nieder}, \&
  {Wieters}}]{sinnhuber2012}
{Sinnhuber}, M., {Nieder}, H., \& {Wieters}, N. 2012, Surveys in Geophysics,
  33, 1281

\bibitem[{Sneep \& Ubachs(2005)}]{sneep2005}
Sneep, M., \& Ubachs, W. 2005, Journal of Quantitative Spectroscopy and
  Radiative Transfer, 92, 293

\bibitem[{Snellen(2014)}]{snellen2014}
Snellen, I. 2014, Philosophical Transactions of the Royal Society A:
  Mathematical, Physical and Engineering Sciences, 372, 20130075.
\newblock
  \url{https://royalsocietypublishing.org/doi/abs/10.1098/rsta.2013.0075}

\bibitem[{Struminsky {et~al.}(2017)Struminsky, Sadovski, \&
  Belov}]{Struminsky2017}
Struminsky, A., Sadovski, A., \& Belov, A. 2017, Cosmic Rays near Proxima
  Centauri b,  Zenodo, {The work was partly supported by the Russian Foundation
  for Basic Research (grant 16-02-00328) and the Programm 1.7 P2 of the Russian
  Academy of Sciences.}, doi:10.5281/zenodo.1064778.
\newblock \url{https://doi.org/10.5281/zenodo.1064778}

\bibitem[{{Tabataba-Vakili} {et~al.}(2016){Tabataba-Vakili}, {Grenfell},
  {Grie{\ss}meier}, \& {Rauer}}]{tabataba2016}
{Tabataba-Vakili}, F., {Grenfell}, J.~L., {Grie{\ss}meier}, J.-M., \& {Rauer},
  H. 2016, \aap, 585, A96

\bibitem[{Th{\'e}bault {et~al.}(2015)Th{\'e}bault, Finlay, Beggan, Alken,
  Aubert, Barrois, Bertrand, Bondar, Boness, Brocco, Canet, Chambodut,
  Chulliat, Co{\"\i}sson, Civet, Du, Fournier, Fratter, Gillet, Hamilton,
  Hamoudi, Hulot, Jager, Korte, Kuang, Lalanne, Langlais, L{\'e}ger, Lesur,
  Lowes, Macmillan, Mandea, Manoj, Maus, Olsen, Petrov, Ridley, Rother, Sabaka,
  Saturnino, Schachtschneider, Sirol, Tangborn, Thomson, T{\o}ffner-Clausen,
  Vigneron, Wardinski, \& Zvereva}]{Thebault2015}
Th{\'e}bault, E., Finlay, C.~C., Beggan, C.~D., {et~al.} 2015, Earth, Planets
  and Space, 67, 79

\bibitem[{{Tilley} {et~al.}(2017){Tilley}, {Segura}, {Meadows}, {Hawley}, \&
  {Davenport}}]{Tilley2017}
{Tilley}, M.~A., {Segura}, A., {Meadows}, V.~S., {Hawley}, S., \& {Davenport},
  J. 2017, ArXiv e-prints, arXiv:1711.08484

\bibitem[{Turbet {et~al.}(2016)Turbet, Leconte, Selsis, Bolmont, Forget, Ribas,
  Raymond, \& Anglada-Escud\'e}]{turbet2016}
Turbet, M., Leconte, J., Selsis, F., {et~al.} 2016, A\&A, 596, A112.
\newblock \url{https://doi.org/10.1051/0004-6361/201629577}

\bibitem[{Valencia {et~al.}(2007)Valencia, Sasselov, \&
  O'Connell}]{valencia2007}
Valencia, D., Sasselov, D.~D., \& O'Connell, R.~J. 2007, The Astrophysical
  Journal, 665, 1413.
\newblock \url{https://doi.org/10.1086%2F519554}

\bibitem[{{Vidotto} {et~al.}(2013){Vidotto}, {Jardine}, {Morin}, {Donati},
  {Lang}, \& {Russell}}]{vidotto2013}
{Vidotto}, A.~A., {Jardine}, M., {Morin}, J., {et~al.} 2013, \aap, 557, A67

\bibitem[{von Paris {et~al.}(2015)von Paris, Selsis, Godolt, Grenfell, Rauer,
  \& Stracke}]{vonParis2015}
von Paris, P., Selsis, F., Godolt, M., {et~al.} 2015, Icarus, 257, 406 .
\newblock
  \url{http://www.sciencedirect.com/science/article/pii/S0019103515002389}

\bibitem[{{Winkler} {et~al.}(2009){Winkler}, {Kazeminejad}, {Sinnhuber},
  {Kallenrode}, \& {Notholt}}]{winkler2009}
{Winkler}, H., {Kazeminejad}, S., {Sinnhuber}, M., {Kallenrode}, M.~B., \&
  {Notholt}, J. 2009, Journal of Geophysical Research (Atmospheres), 114,
  D00I03

\end{thebibliography}



\end{document}